# Efficient tracking of the state of a qubit

Raisa I. Karasik[1], and Howard M. Wiseman[1]

[1]Centre for Quantum Computer Technology, Centre for Quantum Dynamics, Griffith University, Nathan QLD 4111, Australia

*Abstract Summary (35 words)*

*We study resource requirements for tracking the state of an open qubit subject to continuous observation. This problem is of great importance for quantum control.*

*Open quantum systems; continuous measurement; quantum control;*

## I. INTRODUCTION

Tracking the state of an open quantum system with classical resources as it undergoes continuous monitoring is of great interest for quantum control [1-2]. In studying this question, we are faced with puzzling phenomena. A simple system, such as an open qubit, under generic continuous monitoring that resolves every jump, always leaves the system in a pure state and on average yields the same dynamics as unmonitored system[1, 3, 4], exhibits very complex evolution and explores a manifold of pure states. Thus it seems that tracking the state of such system requires infinite memory even though the studied system is just a qubit. We would like to claim that such complex evolution arises from using the 'wrong' measurement scheme and is not intrinsic to the quantum system. Here we study two-level atom undergoing resonance fluorescence. We show that there are up to three different adaptive monitoring schemes that correspond to the simplicity of the system and leave the system jumping between two different states. We also show that different monitoring schemes can be characterized by their efficiency and under certain circumstances can reach very efficient tracking.

## II. RESONANCE FLUORENSCENES

We consider continuous monitoring of the damped and driven two-level atom. The ground state of this atom is denoted by $|g\rangle$ and the excited state is $|e\rangle$. The atom decays to the ground state with rate $\gamma$, but is also driven with a classical field of Rabi frequency $\Omega$. The evolution of the state of the atom is described by the master equation:

$$\dot{\rho} = -i\frac{\Omega}{2}[\sigma_x, \rho] + \frac{\gamma}{2}\left([\sigma\rho, \sigma^\dagger] + [\sigma, \rho\sigma^\dagger]\right).$$

Here density matrix $\rho$ describes the state of the two-level atom, $\sigma = |g\rangle\langle e|$ and $\sigma_x = \sigma + \sigma^\dagger$. This system in the long time limit converges to a mixed steady state

$$\rho_s = \frac{1}{\gamma^2 + 2\Omega^2}\begin{pmatrix} \Omega^2 & -i\gamma\Omega \\ i\gamma\Omega & \gamma^2 + \Omega^2 \end{pmatrix} \quad (1)$$

written in the representation, where $|e\rangle = (1,0)^T$ and $|g\rangle = (0,1)^T$.

To describe continuous measurements of the state of the two-level atom, one might look at the unconditionally infinitesimally evolved density matrix (i.e., different form of the master equation presented above):

$$\rho(t+dt) = \sum_l \Omega_l(dt)\rho(t)\Omega_l^\dagger(dt) \quad (2),$$

where $\Omega_l(dt)$ describes a set of measurements performed on the system. In the simplest possible case, the sum in (2) contains two terms and

$$\Omega_0(dt) = 1 - \left(i\frac{\Omega}{2}\sigma_x + \frac{\gamma}{2}\sigma^\dagger\sigma + \mu^*\gamma\sigma + \frac{\gamma|\mu|^2}{2}\right)dt$$

$$\Omega_1(dt) = \sqrt{\gamma dt}\left(\sigma + \mu\right).$$

From the form of these operators, it is evident that the first gives smooth (no-jump) evolution of infinitesimal duration, while the second gives a quantum jump. Physically, these correspond to not detecting, and detecting, a photon respectively. The quantum jump does not take the atom into the ground state in general because we have allowed for the radiation from the two-level atom to be interfered with a local oscillator of amplitude $\mu$ prior to detection, just as in homodyne detection. Unlike homodyne detection, amplitude of the local oscillator is weak (in comparison to the field from the atom) so that individual photons can be resolved during the detection yielding a jump evolution, not diffusive one observe in homodyne detection. For $\mu = 1/2$ and $\Omega = \gamma = 1$ with initial state $|e\rangle$, conditional evolution of the state of the two-level atom is depicted as trajectories on the Bloch sphere in Fig. 1. Observe that under this generic monitoring, the system evolves between infinitely many pure states. Actually, it covers a manifold of pure states. Such complex evolution might result from the intrinsic complexity of the studied system or might be introduced by the monitoring procedure. Given the simplicity of the two-level system, we expect that simpler evolution is possible with some particular measurement scheme.

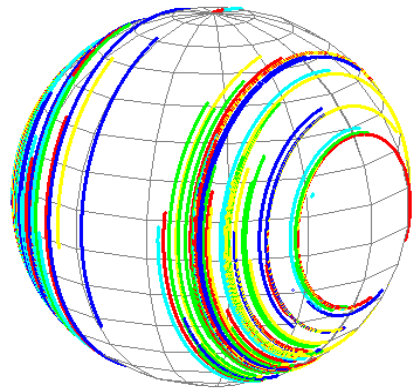

Figure 1. Possible evolutions of the two-level subject to continuous monitoring as trajectories on the Bloch sphere.

### III. TWO-STATE JUMPING

We now search for an adaptive monitoring scheme for the evolution of the two-level atom so that the atom is always in one of the two possible states. In order to solve this problem, we review the analysis developed in [5] and extend it.

Suppose there exists monitoring scheme such that the qubit jumps between $|\psi_1\rangle$ and $|\psi_2\rangle$. This allows us to introduce interpretation for measurement operators from equation (2). For some particular value of $\mu$, operator $\Omega_0$ describes the evolution of the system when the system remains in the same state as before and operator $\Omega_1$ describes the jump to a different state. This description implies that $|\psi_1\rangle$ and $|\psi_2\rangle$ must be eigenstates of $\Omega_0$. In turn, this imposes very strict conditions on $\Omega_0$ and $\Omega_1$, which might be never satisfied. Thus we are unlikely to find a non-adaptive monitoring scheme that always observes atom in one of the two possible states. We, therefore resort, to an adaptive monitoring, which enables us to change a monitoring scheme (i.e., pick a different value for $\mu$) once a jump is detected.

The strategy to address this problem was proposed in [5]. For system in state $|\psi_1\rangle$, we pick $\mu_1$ so that $|\psi_1\rangle$ is an eigenstate of $\Omega_0(\mu_1)$ and $\Omega_1(\mu_1)$ takes $|\psi_1\rangle$ to $|\psi_2\rangle$. Once the jump is detected, the amplitude of the local oscillator is tuned to $\mu_2$ so that $|\psi_2\rangle$ is an eigenstate of $\Omega_0(\mu_2)$ and $\Omega_1(\mu_2)$ takes $|\psi_1\rangle$ to $|\psi_2\rangle$. Observe that this scheme is only possible if such $\mu_1$ and $\mu_2$ exist.

Unlike [5], we only interested in any such states $|\psi_1\rangle$ and $|\psi_2\rangle$ that fulfill

$$p_1|\psi_1\rangle\langle\psi_1| + p_2|\psi_2\rangle\langle\psi_2| = \rho_s \quad (3)$$

with $p_1 + p_2 = 1$ to ensure that resulting dynamics reproduces the average of the unmonitored dynamics. Here $p_1$ represents the proportion (probability) the system spends in $|\psi_1\rangle$ and $p_2$ is the probability for the system to be in $|\psi_2\rangle$. The necessity for this condition was shown in [6].

#### A. Determining the adaptive scheme

In this section, we determine parameters $\mu_1$ and $\mu_2$ for the adaptive monitoring that always takes the atom between two states. The system that jumps between two different states, returns to the original states after two jumps, i.e., $\Omega_1(\mu_2)\Omega_1(\mu_1)|\psi_i\rangle \propto |\psi_i\rangle$. Recalling the definition for $\Omega_1$, we get

$$(\sigma + \mu_2)(\sigma + \mu_1)|\psi_i\rangle \propto |\psi_i\rangle. \quad (4)$$

i.e., $|\psi_i\rangle$ is an eigenstate of

$$\begin{pmatrix} \mu_1\mu_2 & 0 \\ \mu_1 + \mu_2 & \mu_1\mu_2 \end{pmatrix}. \quad (5)$$

which has only one eigenvector (no two-state jump solution) unless

$$\mu_1 = -\mu_2. \quad (6)$$

If, on the other hand, (6) holds, then operator in (5) is proportional to the identity matrix. Any state is an eigenstate. Thus no additional conditions are imposed.

Also states $|\psi_1\rangle$ and $|\psi_2\rangle$ must be stationary with respect to no-jump evolution $\Omega_0$ from (1), i.e., $|\psi_1\rangle$ and $|\psi_2\rangle$ must be respective eigenstates of $\Omega_0(\mu_1)$ and $\Omega_0(\mu_2)$. Eigenstates of $\Omega_0(\mu)$ take the form

$$|\psi^\mu_{\pm e}\rangle = \left(\tfrac{i\gamma}{2} \pm f(\mu)\right)|g\rangle + \Omega|e\rangle \quad (7)$$

with $f(\mu) = \sqrt{\Omega^2 - 2i\Omega\gamma\mu^* - \gamma^2/4}$.

Equation (7) allows us to determine the value of $\mu_1$. Suppose we start from the state $|\psi^{\mu_1}_{+e}\rangle$. Then after the first jump the atom will be in state $(\sigma + \mu_1)|\psi^{\mu_1}_{+e}\rangle$, which must be an eigenstate of $\Omega_0(\mu_2) = \Omega_0(-\mu_1)$. With the help of (6) and (7), this requirement reduces to

$$f(\gamma, \Omega, -\mu_1) = f(\gamma, \Omega, \mu_1) - \tfrac{\Omega}{\mu_1} \quad (8)$$

Equivalent condition can be derived if the evolution starts with $|\psi^{\mu_1}_{-e}\rangle$.

From now on, we denote $\mu_1$ by $\mu$. We also convert (8) to the radical-free form

$$4\gamma^2 |\mu|^4 \mu^2 + (\Omega^2 - \tfrac{\gamma^2}{4})\mu^2 = \tfrac{\Omega^2}{4}. \quad (9)$$

To solve this equation, we decompose $\mu$ into real and imaginary parts ($\mu = \mu_r + i\mu_i$). Note that in this decomposition $\mu_r$ and $\mu_i$ are real. We also introduce the function $G = 4\gamma^2(\mu_r^2 + \mu_i^2)^2 + (\Omega^2 - \gamma^2/4)$. Then the real part of (9) is

$$G(\mu_r^2 - \mu_i^2) - \tfrac{\Omega^2}{4} = 0 \quad (10)$$

and imaginary part of (7) is

$$2\mu_r \mu_i G = 0. \quad (11)$$

We solve (10-11) for $\mu_r$ and $\mu_i$. Equation (11) is zero if $\mu_r = 0$ or $\mu_i = 0$ or $G = 0$. The last condition cannot be satisfied together with (10).

If $\mu_i = 0$, then (10) reduces to

$$0 = 4\gamma^2 \mu_r^6 + (\Omega^2 - \tfrac{\gamma^2}{4})\mu_r^2 - \tfrac{\Omega^2}{4}$$

$$= \left(\mu_r^2 - \tfrac{1}{4}\right)\left(4\gamma^2\left(\mu_r^4 + \tfrac{\mu_r^2}{4}\right) + \Omega^2\right)$$

For $\Omega \neq 0$, the second factor in the above equation is always positive and, therefore, (10) equals to zero for $\mu_r = \pm\tfrac{1}{2}$.

If $\mu_r = 0$, then (10) reduces to

$$0 = 4\gamma^2 \mu_i^6 + (\Omega^2 - \tfrac{\gamma^2}{4})\mu_i^2 + \tfrac{\Omega^2}{4}$$

$$= \left(\mu_i^2 + \tfrac{1}{4}\right)\left(4\gamma^2\left(\mu_i^4 - \tfrac{\mu_i^2}{4}\right) + \Omega^2\right)$$

Here the first factor is always positive, but the second factor yields a solution for

$$\mu_i = \pm\sqrt{\tfrac{1}{8} \pm \tfrac{1}{8}\sqrt{1 - 16\tfrac{\Omega^2}{\gamma^2}}}.$$

By construction $\mu_i$ must be real. Therefore this solution exists only when $\Omega \leq \gamma/4$.

At this point, we have identified all possible values for $\mu$ that allow for two-state jumping for two-level atom undergoing resonance fluorescence. For each $\mu$, using (7) we can identify four potential pairs for two state jumping:

$$\left\{\left|\psi_{+e,+\mu}\right\rangle, \left|\psi_{+e,-\mu}\right\rangle\right\}, \left\{\left|\psi_{+e,+\mu}\right\rangle, \left|\psi_{-e,-\mu}\right\rangle\right\},$$

$$\left\{\left|\psi_{-e,+\mu}\right\rangle, \left|\psi_{-e,-\mu}\right\rangle\right\}, \left\{\left|\psi_{-e,+\mu}\right\rangle, \left|\psi_{+e,-\mu}\right\rangle\right\}.$$

However, not all of them satisfy (3) and, thus, will not be consistent with the average of the unmonitored evolution and are not good for tracking.

### B. Identify the two-state jumping solutions and computing the entropy

For each possible value of $\mu_1$, we look at 4 possible arrangements and numerically identify a pair that satisfies equation (3).

As shown in [5], for $\mu_1 = \tfrac{1}{2}$, $\left\{\left|\psi_{-e,+\mu}\right\rangle, \left|\psi_{+e,-\mu}\right\rangle\right\}$ describes two-state jumping and the probability to occupy each state is $\tfrac{1}{2}$.

For $\mu_1 = i\sqrt{1 + \sqrt{1 - 16\Omega^2/\gamma^2}}/\sqrt{8}$, the situation is much more complex. For $|\Omega| \leq 0.242\gamma$ the states $\left\{\left|\psi_{+e,+\mu}\right\rangle, \left|\psi_{+e,-\mu}\right\rangle\right\}$ give rise to two-state jumping. In the region $0.25\gamma \geq \Omega > 0.242\gamma$, the solutions are given by $\left\{\left|\psi_{+e,+\mu}\right\rangle, \left|\psi_{-e,-\mu}\right\rangle\right\}$ and $\left\{\left|\psi_{-e,+\mu}\right\rangle, \left|\psi_{+e,-\mu}\right\rangle\right\}$ are solutions when $-0.25\gamma \leq \Omega < -0.242\gamma$. In all of these cases, probabilities for occupying each states are not equal.

For $\mu_1 = i\sqrt{1 - \sqrt{1 - 16\Omega^2/\gamma^2}}/\sqrt{8}$, system jumps between set of states $\left\{\left|\psi_{+e,+\mu}\right\rangle, \left|\psi_{-e,-\mu}\right\rangle\right\}$ when $0.25\gamma \geq \Omega > 0$ and $\left\{\left|\psi_{-e,+\mu}\right\rangle, \left|\psi_{+e,-\mu}\right\rangle\right\}$ when $-0.25\gamma \leq \Omega < 0$. Again probabilities for occupying each states are not equal.

Under continuous monitoring the atom will occupy states $|\psi_1\rangle$ and $|\psi_2\rangle$ with respective probabilities $p_1$ and $p_2$. We can quantify the uncertainty with occupying states $|\psi_1\rangle$ and $|\psi_2\rangle$ with the Shannon entropy,

$$h[\{p_1, p_2\}] = -p_1 \log_2 p_1 - p_2 \log_2 p_2.$$

The Shannon entropy quantifies amount of information contained in this ensemble and determines the necessary number of bits for lossless data compression. We compute the Shannon entropy for each ensemble we have identified and plot it figure 2. Solutions with varied $\mu$ have entropy less than one. This means that an ensemble with a large number $N$ of qubits, each independently monitored, can be tracked with just $hN$ bits by using data compression on the classical memory. Solutions that arise when

$$\mu_1 = i\sqrt{1-\sqrt{1-16\Omega^2/\gamma^2}}\Big/\sqrt{8},$$

can be tracked in the most efficient manner.

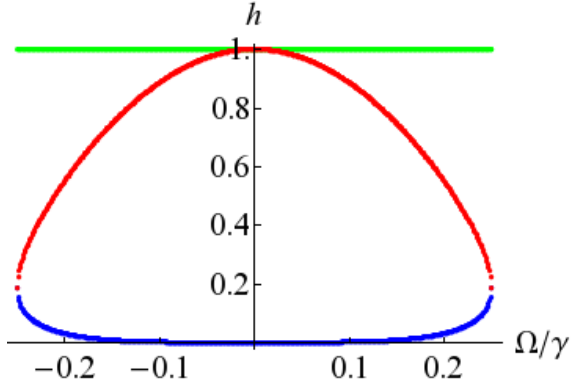

Figure 2: Entropy curves associated with three different two-state jumping solutions.

In summary, we have studied two-level atom undergoing resonance fluorescence and subject to continuous monitoring. Generic continuous monitoring gives very complex evolution, where two-level atom jumps between manifold of pure states. In this paper, we found non-adaptive monitoring that is consistent with the simplicity of the system and it is just jumping between two non-orthogonal states. These states can be used for efficient tracking of the state of the system.

We have since shown, in Ref. [7], that it is always possible to find at least one such pair of non-orthogonal states, for any master equation for a qubit. We also presented evidence that for higher-dimensional quantum systems, such finite ensembles of pure states also exist, although the number of states is expected to scale with the square of the Hilbert space dimension in general. However, it is an open problem as to how to generalize the method used in this paper to find explicitly the monitoring scheme that will physically realize a given finite ensemble that is known to be physically realizable.

ACKNOWLEDGMENT

This work was funded by the ARC grants CE0348250 and FF0458313.

REFERENCES

[1] H. M. Wiseman and G. J. Milburn, "Quantum Measurement and Control", Cambridge University Press, (2009).
[2] Hideo Mabuchi, "Coherent-feedback quantum control with a dynamic compensator," Phys. Rev. A **78**, 032323, (2008).
[3] H. J. Carmichael, "An Open Systems Approach to Quantum Optics", Springer-Verlag, (1993).
[4] H. M. Wiseman, "Quantum trajectories and quantum measurement theory", Quantum Semiclass. Opt. **8**, 205, (1996).
[5] H. M. Wiseman and G. E. Toombes, "Quantum jumps in a two-level atom: simples theories versus quantum trajectories", Phys. Rev. A, **60**, 2474, (1999).
[6] H. M. Wiseman and J. A. Vaccaro, "Inequivalence of pure state ensembles for open quantum systems: the preferred ensembles are those that are physically realizable", Phys. Rev. Lett., 87, 240402 (2001).
[7] R. I. Karasik and H. M. Wiseman, "How many bits does it take to track an open quantum system?", arXiv:1009.2675, (2010).